\documentstyle[12pt,fleqn]{article}
\input epsf

\begin{document}
\begin{titlepage}
\vspace*{-60pt}
\begin{flushright}
{\small 
SUSX-TH-98-001 \\
SUSSEX-AST 98/2-4 \\
FERMILAB-Pub-97/434-A \\
astro-ph/9802209}\\
\end{flushright}
\vspace*{10pt}
\begin{center}
\Large {\bf On the reliability of inflaton \\ potential reconstruction}
\vspace{.8cm}
\normalsize

{\bf Edmund J.\ Copeland} \\
{\it Centre for Theoretical Physics, University of Sussex\\ Falmer,  
Brighton BN1 9QH, United Kingdom} \\
\vspace{.4 cm}
{\bf Ian J.\ Grivell} \\
{\it Astronomy Centre, University of Sussex \\Falmer, Brighton BN1  
9QJ, United Kingdom} \\ 
\vspace{.4 cm}
{\bf Edward W.\ Kolb} \\
{\it NASA/Fermilab Astrophysics Center, Fermi National Accelerator 
Laboratory\\ Batavia, Illinois 60510\\
and \\
Department of Astronomy and Astrophysics and Enrico Fermi Institute\\ 
The University of Chicago, Chicago, Illinois 60637} \\
\vspace{.4 cm}
{\bf Andrew R.\ Liddle} \\
{\it Astronomy Centre, University of Sussex\\ Falmer, Brighton BN1  
9QJ, United Kingdom} 
\date{\today} 
\end{center}
\begin{abstract} 
If primordial scalar and tensor perturbation spectra can be inferred
from observations of the cosmic background radiation and large-scale
structure, then one might hope to reconstruct a unique single-field
inflaton potential capable of generating the observed spectra.  In
this paper we examine conditions under which such a potential can be
reliably reconstructed. For it to be possible at all, the spectra must
be well fit by a Taylor series expansion. A complete reconstruction
requires a statistically-significant tensor mode to be measured in the
microwave background. We find that the observational uncertainties
dominate the theoretical error from use of the slow-roll
approximation, and conclude that the reconstruction procedure will
never insidiously lead to an irrelevant potential.

\vspace{24pt}
\end{abstract}
\end{titlepage}

\section{Inflation and Perturbations}

Inflation produces metric perturbations, which are presently the most
plausible cause of the observed temperature fluctuations in the cosmic
background radiation (CBR) and which may act as the seeds for
structure formation.  During inflation the Hubble radius $H^{-1}$
($H=\dot{a}/a$ is the expansion rate) must have increased more slowly
than the scale factor $a$, while during the radiation and matter eras
it increased as $a^2$ and $a^{3/2}$, respectively.  The dynamics of
the evolution of the Hubble radius (or, equivalently, the evolution of
the expansion rate $H$) during inflation is usually modelled by
assuming that the energy density is dominated by scalar-field
potential energy.  The number of dynamical degrees of freedom
associated with the evolution of the scalar-field potential energy
(and hence the Hubble radius) is unknown.  If there is only one
relevant degree of freedom, that degree of freedom is called the {\it
inflaton,} and the scalar-field potential is called the {\it inflaton
potential.}

Inflation models where a single inflaton field slowly evolves under
the influence of the inflaton potential are called single-field
slow-roll models.  Such models are very attractive because of their
simplicity, because they arise naturally in a host of particle physics
models, and because other more complicated models can often be
expressed in terms of an effective single-field slow-roll model.

The assumption that only a single scalar field is dynamically relevant
can, at least in principle, be tested by a series of consistency
relations which represent relations between the scalar and tensor
perturbations produced in single-field models \cite{Recon,LLKCBA}.
Inflation models where more than one scalar field is dynamically
important are phenomenologically much more complicated, and almost
certainly must be studied on a case-by-case basis.  In this paper we
shall restrict our discussion to the single-field case, where the
scalar potential $V(\phi)$ is the only piece of information to be
specified.

Reconstruction of the inflaton potential (see Ref.\ \cite{LLKCBA} for
a review) refers to the process of using observational data,
especially microwave background anisotropies, to determine the
inflaton potential capable of generating the perturbation spectra
inferred from observations \cite{Recon}.  Of course there is no way to
prove that the reconstructed inflaton potential was the agent
responsible for generating the perturbations.  What can be hoped for
is that one can determine a {\it unique} (within observational errors)
inflaton potential capable of producing the observed perturbation
spectra.  The reconstructed inflaton potential may well be the first
concrete piece of information to be obtained about physics at scales
close to the Planck scale.

As is well known, inflation produces both scalar and tensor
perturbations, which each generate microwave anisotropies (see Ref.\
\cite{LLrep} for a review).  As reviewed in the next section, if
$V(\phi)$ is given  the perturbation spectra can be computed exactly
in linear perturbation theory through integration of the relevant mode
equations.  If the scalar field is rolling sufficiently slowly, the
solutions to the mode equations may be approximated using the
slow-roll expansion \cite{SB,LL,LPB}. The standard reconstruction
program makes use of the slow-roll expansion, taking advantage of a
calculation of the perturbation spectra by Stewart and Lyth \cite{SL},
which gives the next-order correction to the usual lowest-order
slow-roll results.

The biggest hurdle for successful reconstruction is that many
inflation models predict a tensor perturbation amplitude well below
the expected threshold for detection, and as we will see, even above
detection threshold the errors can be considerable. If the tensor
modes cannot be identified a unique reconstruction is impossible, as
the scalar perturbations are governed not only by $V(\phi)$, but by
the first derivative of $V(\phi)$ as well.  Knowledge of only the
scalar perturbations leaves an undetermined integration constant in
the non-linear system of reconstruction equations.  Another problem is
that simple potentials usually lead to a nearly exact power-law scalar
spectrum with the spectral index close to unity.  In such a scenario
only a very limited amount of information could be obtained about high
energy physics from astrophysical observations.

Recently, Wang et al.\ \cite{WMS} suggested an alternative danger to
reconstruction.  They studied the possibility that the potential might
be sufficiently complicated that the slow-roll expansion breaks down,
and a full integration of the mode functions is necessary to calculate
the perturbation spectra to the accuracy expected by the next round of
satellite observations by {\sc map} and Planck.

It is, of course, clear that any perturbative procedure such as the
slow-roll expansion has some limit to its validity.  But here, the
breakdown of the slow-roll expansion is particularly troublesome;
firstly because of the technical difficulty that there is no known
method of computing yet higher-order corrections, and secondly because
there is a general possibility that for sufficiently complicated
potentials the perturbation series may not converge at all. However,
the key concern for reconstruction is whether such a situation will
lead to a {\it misconstruction} of the inflaton potential.

In this paper, we shall argue that for the purpose of reconstruction
the breakdown of the slow-roll approximation is in many senses
desirable. First of all, we show that there is no practical danger
that one can be misled into thinking perturbative reconstruction is
working when in fact it is not. Secondly, the very failure of the
perturbative approach implies that extra parameters are required in
order to explain the observed data set, and hence there is the
prospect of obtaining a greater amount of information about the
inflationary model. Finally, we will demonstrate that the theoretical
errors from the use of slow-roll equations are sub-dominant to the
expected observational errors from a satellite experiment such as
Planck.

Wang et al.\ \cite{WMS} described a particular, admittedly rather {\em
ad hoc}, scalar field potential, for which they demonstrated that the
result of exact mode equation integration differed significantly, by
the standards of say the Planck satellite, from the slow-roll
expansion. We shall largely concentrate on that potential as a means
of concrete illustration of the very general points we shall make.

\section{The slow-roll expansion}
\setcounter{equation}{0}

Two crucial ingredients for reconstruction are the primordial scalar and 
tensor perturbation spectra $A_S(k)$ and $A_T(k)$ which we define as in 
Ref.~\cite{LLKCBA}. 
We define slow-roll parameters $\epsilon$, $\eta$, and $\xi$ as 
\begin{eqnarray}
\label{epsilon}
\epsilon (\phi) & \equiv & \frac{m_{{\rm Pl}}^2}{4\pi} \left[ 
	\frac{H' (\phi) }{H(\phi)} \right]^2 \\
\label{eta}
\eta (\phi)  & \equiv &  \frac{m_{{\rm Pl}}^2}{4\pi}  
	\frac{H'' (\phi) }{H(\phi)}  \\
\label{xi}
\xi (\phi) & \equiv &  \frac{m_{{\rm Pl}}^2}{4\pi} \left( \frac{H'(\phi) H''' 
	(\phi)}{H^2(\phi)} \right)^{1/2} \ ,
\end{eqnarray}
where here and throughout
the paper the prime superscript implies $d/d\phi$. As long as these 
parameters are small compared to unity, 
$A_S(k)$ and $A_T(k)$ are given by \cite{SL} 
\begin{eqnarray}
\label{secondscalar}
A_S^2(k) & \simeq & \frac{4}{25 \pi}\left(\frac{H}{m_{{\rm Pl}}} \right)^2 
\left(\frac{1}{\epsilon}\right) \left[1-(2C+1) \epsilon + C\eta \right]^2  \\
\label{secondtensor}
A_T^2(k) & \simeq & \frac{4}{25 \pi}\left(\frac{H}{m_{{\rm Pl}}} \right)^2
\left[ 1-(C+1 ) \epsilon \right] ^2\ ,
\end{eqnarray}
where $H(\phi)$, $\epsilon(\phi)$, and $\eta(\phi)$ are to be
determined at the value of $\phi$ when $k=aH$ during inflation, and
where $C = -2 +\ln 2 + \gamma \simeq -0.73$ is a numerical constant,
$\gamma$ being the Euler constant.  These equations are the basis of
the reconstruction process and provide an analytic alternative to 
integration of the full mode equations. 

The expressions for $A^2_S(k)$ and $A^2_T(k)$ are in terms of
$H(\phi)$ and the slow-roll parameters $\epsilon(\phi)$ and
$\eta(\phi)$, but what is usually known is the potential $V(\phi)$.
Since $H$ can be written exactly in terms of $V$ and $\epsilon$:
\begin{equation}
\label{heps}
\frac{H^2(\phi)}{m_{{\rm Pl}}^2} = 8\pi\,\frac{V(\phi)}{m_{{\rm Pl}}^4}\,
\frac{1}{3-\epsilon(\phi)} \ ,
\end{equation}
the problem reduces to finding $\epsilon$ and $\eta$ in terms of $V$
and its derivatives.

There are two methods to determine $H$ and the slow-roll parameters.
The first is an analytic approach studied by Kolb and Vadas \cite{KV}
and by Liddle et al.\ \cite{LPB}.  (The results are identical; here we
follow the notation of the later paper.)  The approach involves an
infinite hierarchy of potential slow-roll parameters:
\begin{eqnarray}
\label{epvd}
\epsilon_{\scriptscriptstyle V}(\phi)  =  
\frac{m_{{\rm Pl}}^2}{16\pi} \left( \frac{V'}{V} \right)^2 \qquad & ; & 
\qquad
	\eta_{\scriptscriptstyle V}(\phi) = \frac{m_{{\rm Pl}}^2}{8\pi}\, 
	\frac{V''}{V} \,, \nonumber \\
\xi^2_{{\scriptscriptstyle V}} (\phi)  = 
	\frac{m^{2}_{{\rm Pl}}}{8\pi}\,
	\frac{\left(V'\right)^1V^{(3)}}{V^{2}} 
	\qquad & ; & \qquad \sigma^3_{{\scriptscriptstyle V}}(\phi) = 
	\frac{m^{2}_{{\rm Pl}}}{8\pi}\,
	\frac{\left(V'\right)^{2}V^{(4)}}{V^{3}}\,, 
\end{eqnarray}
and so on, where $V^{(n)}=d^nV/d\phi^n$.

The slow-roll parameters of Eqs.\ (\ref{epsilon}-\ref{xi}) involve an
infinite number of potential slow-roll parameters, and to obtain an
analytic expression it is necessary to truncate the expression by
neglecting potential slow-roll parameters above a given order.  For
most potentials the higher derivatives become small rapidly, and the
expressions can be terminated after only a few terms.  For instance, to
fourth order in $H^{2}$ \cite{LPB}
\begin{eqnarray}
\label{semih}
H^{2}(\phi) & = & \frac{8\pi}{3m_{{\rm Pl}}^2}V(\phi)\left[1 
+ \frac{1}{3}\epsilon_{{\scriptscriptstyle V}} 
- \frac{1}{3}\epsilon^{2}_{{\scriptscriptstyle V}} +
+ \frac{2}{9}\epsilon_{{\scriptscriptstyle V}}
        \eta_{{\scriptscriptstyle V}} + 
\frac{25}{27}\epsilon^{3}_{{\scriptscriptstyle V}} 
\right. \nonumber \\ & &  
+ \frac{5}{27}\epsilon_{{\scriptscriptstyle V}}
           \eta^{2}_{{\scriptscriptstyle V}} 
 - \frac{26}{27}\epsilon^{2}_{{\scriptscriptstyle V}}
          \eta_{{\scriptscriptstyle V}}  
+ \frac{2}{27}\epsilon_{{\scriptscriptstyle V}}
          \xi^{2}_{{\scriptscriptstyle V}} - 
\frac{327}{81}\epsilon^{4}_{{\scriptscriptstyle V}}
+ \frac{460}{81}\epsilon^{3}_{{\scriptscriptstyle V}}
         \eta_{{\scriptscriptstyle V}} - 
\frac{172}{81}\epsilon^{2}_{{\scriptscriptstyle V}}
                  \eta^{2}_{{\scriptscriptstyle V}} 
\nonumber \\ & &
+ \left.\frac{14}{81}\epsilon_{{\scriptscriptstyle V}}
            \eta^{3}_{{\scriptscriptstyle V}}
- \frac{44}{81}\epsilon^{2}_{{\scriptscriptstyle V}}
             \xi^{2}_{{\scriptscriptstyle V}} 
+ \frac{2}{9}\epsilon_{{\scriptscriptstyle V}}
                    \eta_{{\scriptscriptstyle V}}
              \xi^{2}_{{\scriptscriptstyle V}}
 + \frac{2}{81}\epsilon_{{\scriptscriptstyle V}}
              \sigma^{3}_{{\scriptscriptstyle V}} + \cdots \, \right]  
\end{eqnarray}
and similar expressions can be found in Refs.~\cite{LPB,KV} for 
$\epsilon(\phi)$, $\eta(\phi)$, etc.

In practice, one would calculate the first few potential slow-roll
parameters, and if the ones involving higher derivatives become
small, the series may be truncated.  However if the higher-derivative
slow-roll parameters do not become small, then it is necessary to
perform a numerical calculation.

The slow-roll parameters may be calculated numerically quite easily by
expressing Eq.\ (\ref{heps}) as an equation for $\epsilon$:
\begin{equation}
\epsilon(\phi) = 3 - 8\pi \frac{V(\phi)}{m^2_{{\rm Pl}}H^2(\phi)} \ ,
\end{equation}
taking the derivative with respect to $\phi$, using the expressions
for $H'$ and $H$ in terms of $\epsilon$ [Eq.\ (\ref{epsilon}) and Eq.\
(\ref{heps})] to yield a first-order differential equation satisfied
by $\epsilon$,
\begin{equation}
\label{epsn}
\epsilon' = \left(\frac{m_{{\rm Pl}} V'}{V} +
\sqrt{16\pi\epsilon}\right)(\epsilon-3).
\end{equation}
It is easy to show that
$\eta=\epsilon-m_{{\rm Pl}}\, \epsilon'/\sqrt{16\pi\epsilon}$, so using Eq.\
(\ref{epsn}) for $\epsilon'$ and integrating it to find $\epsilon$,
$\eta$ is found.  The slow-roll parameter $\xi$ can be expressed in
terms of $\epsilon$, $\eta$, and $\eta'$ as $\xi^2 = \epsilon\eta -
m_{{\rm Pl}}\eta' \sqrt{\epsilon/4\pi}$.  For this expression the additional
function $\epsilon''$ is required, which can be expressed as
\begin{equation}
m^2_{{\rm Pl}}\epsilon'' = \left[ \frac{m^2_{{\rm Pl}}V''}{V}
-\left(\frac{m_{{\rm Pl}}V'}{V}\right)^2 +\frac{8\pi
m_{{\rm Pl}}\,\epsilon'}{\sqrt{16\pi\epsilon}} \right] (\epsilon-3) 
+ \left[ \frac{m_{{\rm Pl}}V'}{V} 
+\sqrt{16\pi\epsilon}\right] .
\end{equation}

\begin{table}
\begin{center}
\begin{tabular}{r|l|l}
\hline \hline
method & $H$; $\epsilon$; $\eta$& $A^2_S(k)$ and  $A^2_T(k)$ \\ 
\hline \hline
numerical & numerical: Eq.\ (\ref{heps}) & mode equations 
        \\ \hline 
semi-analytic &numerical: Eq.\ (\ref{heps}) & slow-roll expressions 
        [Eqs.\ (\ref{secondscalar}) \& (\ref{secondtensor})] \\ \hline
higher-order & $H$:  Eq.\ (\ref{semih}) & \\
analytic & $\epsilon$: Refs.~\cite{LPB,KV} & slow-roll expressions 
                 [Eqs.\ (\ref{secondscalar}) \&(\ref{secondtensor})]\\ 
  & $\eta$: Refs.~\cite{LPB,KV} & \\ \hline
lowest-order & $H=8\pi V/3m^2_{{\rm Pl}}$ & truncation of slow-roll
	expressions\\
analytic & $\epsilon = \epsilon_V$ & 
	\quad \quad [Eqs.\ (\ref{secondscalar}) \& (\ref{secondtensor})] \\ 
	& $\eta=\epsilon_V-\eta_V$ & \\ 
\hline\hline
\end{tabular}
\caption{Different approaches to computing the power spectrum from a 
potential $V(\phi)$.}
\end{center}
\end{table}

For the calculation of $A_S(k)$ and $A_T(k)$ there are three levels of
ease/accuracy one may employ as indicated in Table 1.  The simplest,
but least accurate, method is an analytic approach, using Eq.\
(\ref{semih}) and corresponding ones for $H$, $\epsilon$, and $\eta$,
and the slow-roll expressions, Eqs.\ (\ref{secondscalar}) and
(\ref{secondtensor}), for $A^2_S(k)$ and $A^2_T(k)$.  This will be
accurate if the potential slow-roll parameters
$\epsilon_{{\scriptscriptstyle V}}$, $\eta_{{\scriptscriptstyle V}}$,
etc.~become small, so that a truncation can be performed in the
expressions (the analogues of Eq.\ (\ref{semih})) for the slow-roll
parameters $\epsilon$ and $\eta$, and provided the slow-roll
parameters are smaller than unity.  If $\epsilon$ and $\eta$ are small
enough, the square brackets in Eqs.\ (\ref{secondscalar}) and
(\ref{secondtensor}) may be replaced by unity and the expressions
simplify further.

The most accurate and reliable, but most involved, is a numerical
calculation in which the slow-roll parameters are calculated
numerically and used in a numerical integration of the mode equations.

For some potentials, it may be desirable to perform something we call
a semi-analytic calculation, which involves calculating $\epsilon$ and
$\eta$ numerically, and using the numerical values in the slow-roll
expression for $A^2_S(k)$ and $A^2_T(k)$.  This will be useful if the
higher-order potential slow-roll parameters become large but the
actual values of $\epsilon$, $\eta$, and $\xi$ are small. Only one
numerical integration is required to compute the spectrum, whereas the
mode equation integration must be done for each $k$.

\section{The potential}
\setcounter{equation}{0}

\begin{figure}[t]
\centering 
\leavevmode\epsfysize=8cm \epsfbox{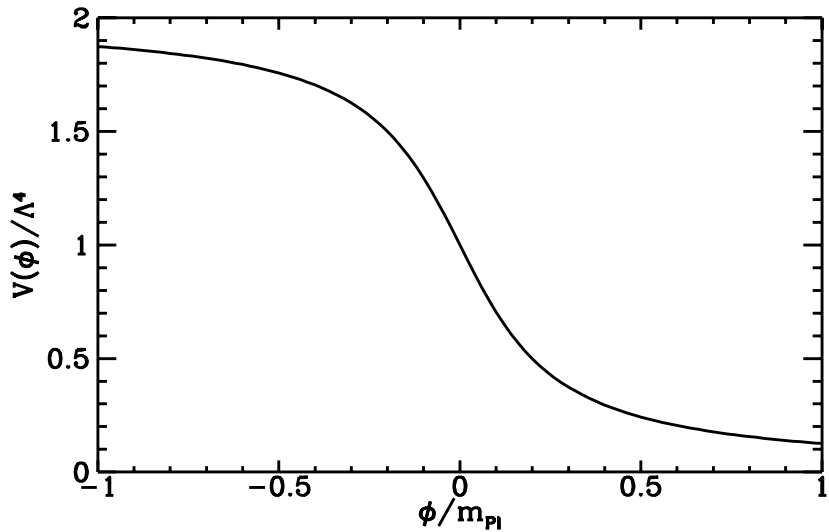}
\caption[fig1]{\label{potential}  The potential of Eq.\ \ref{Vphi}.}
\end{figure} 

Wang et al.\ \cite{WMS} consider an inflaton potential of the form
\begin{equation}
\label{Vphi}
V(\phi) = \Lambda^4 \left( 1 - \frac{2}{\pi} \, \tan^{-1}
	\frac{5\phi}{m_{{\rm Pl}}} \right) \,,
\end{equation}
which is shown in Fig.\ \ref{potential}. We shall take $\Lambda = 1$;
results could be rescaled if desired to match the COBE
normalization. With this potential, inflation is everlasting. Wang et
al.\ \cite{WMS} envisage extra physics, such as a hybrid inflation
mechanism \cite{hybrid}, intervening to end inflation at some point,
and assume that the field's location at the time observable
perturbations were formed can be placed anywhere on the potential. As
a specific example, they choose $\phi=-0.3 m_{{\rm Pl}}$ to be the
value of the field when the present Hubble radius equalled the Hubble
radius during inflation.  We adopt their choice for the end of
inflation.

\begin{figure}[t]
\centering 
\leavevmode\epsfysize=8cm \epsfbox{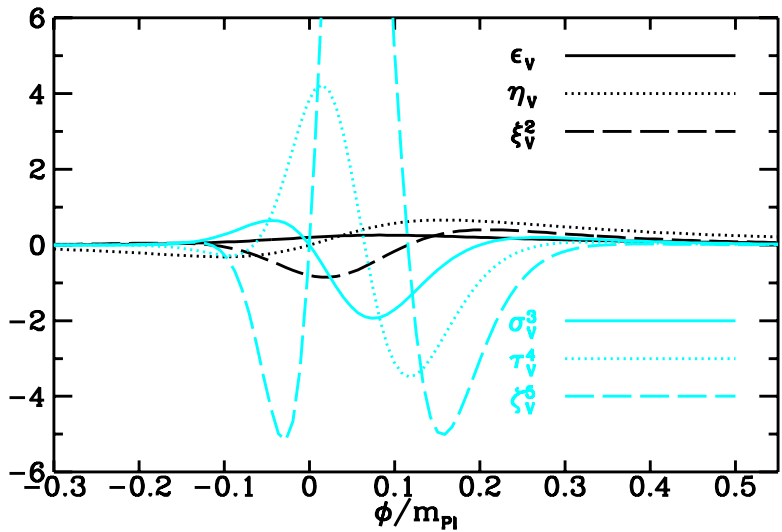}
\caption[fig2]{\label{slow_roll}  The first six slow-roll parameters.}
\end{figure}

Fig.\ \ref{slow_roll} shows the potential slow-roll parameters of Eq.\
(\ref{epvd}), $\tau_{{\scriptscriptstyle V}}$ and
$\zeta_{{\scriptscriptstyle V}}$ being the next two in the hierarchy.
Since potential slow-roll parameters become large, the truncation
necessary to calculate the slow-roll parameters in terms of the
potential slow-roll parameters is not reliable in the vicinity of the
origin, though they have chosen the potential carefully so that the
violation is not sufficient to even temporarily end inflation by
driving $\epsilon$ larger than unity.

\begin{figure}[t]
\centering 
\leavevmode\epsfysize=8cm \epsfbox{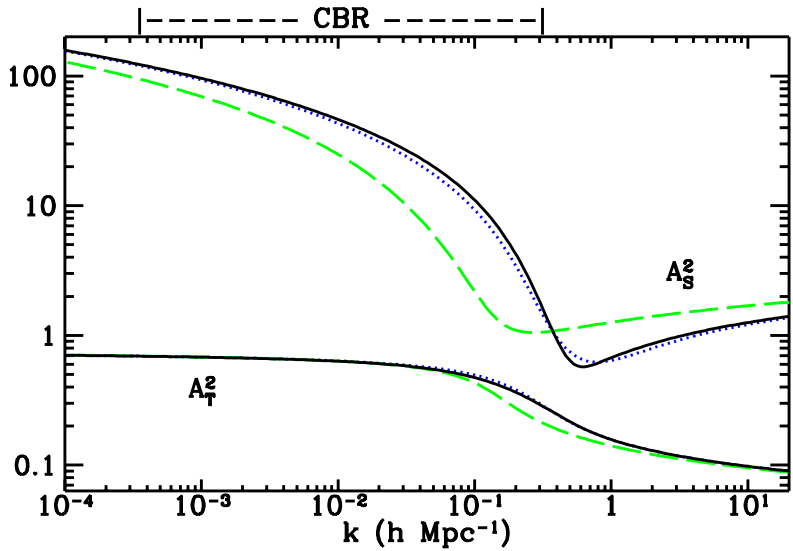}\\ 
\caption[fig3]{\label{asat} The perturbation spectra obtained by mode
equation integration (solid curves) compared to the semi-analytic
results (dotted curves) and the lowest-order analytic results (dashed
curves).  Since the amplitude of the power spectra are proportional to
the unknown parameter $\Lambda$ in the potential, $A_S^2$ and $A_T^2$
may be adjusted up and down by a common factor.  Since inflation is
terminated by hand, the spectra can also be shifted left or
right. Shown in the figure are the spectra analyzed in subsection 5.1.
The spectra analyzed in subsection 5.2 are shifted to the left by a
factor of 30 so that the dip in $A_S^2$ is near $k=0.02h$\,Mpc$^{-1}$.
}
\end{figure}

Given the approximation of linear perturbation theory, the
perturbation spectrum can be computed exactly taking the inflationary
dynamics into account by using the mode equation formalism of Mukhanov
\cite{Muk}.  Numerical solution of this equation was described by
Grivell and Liddle \cite{GL} and Wang et al.\ \cite{WMS}. The scalar
and tensor perturbations generated by this potential are shown in
Fig.\ \ref{asat}. We also show the result of the best available
semi-analytic technique as described above, which uses numerical
solution of the background field equations as input to the
Stewart--Lyth formula and which gives an error of more than ten
percent at some $k$. We see that the matter power spectrum is very far
from the scale-invariant form, featuring a sharp downturn and then
bending back up again. In fact, this deviation from scale-invariance
is already too large to be permitted by observational data; for
example, in a COBE-normalized CDM-like cosmology the perturbations at
$8 h^{-1}$ Mpc are only $\sigma_8 \simeq 0.3$. It still serves as a
useful example however.

\begin{figure}[t]
\centering 
\leavevmode\epsfysize=8cm \epsfbox{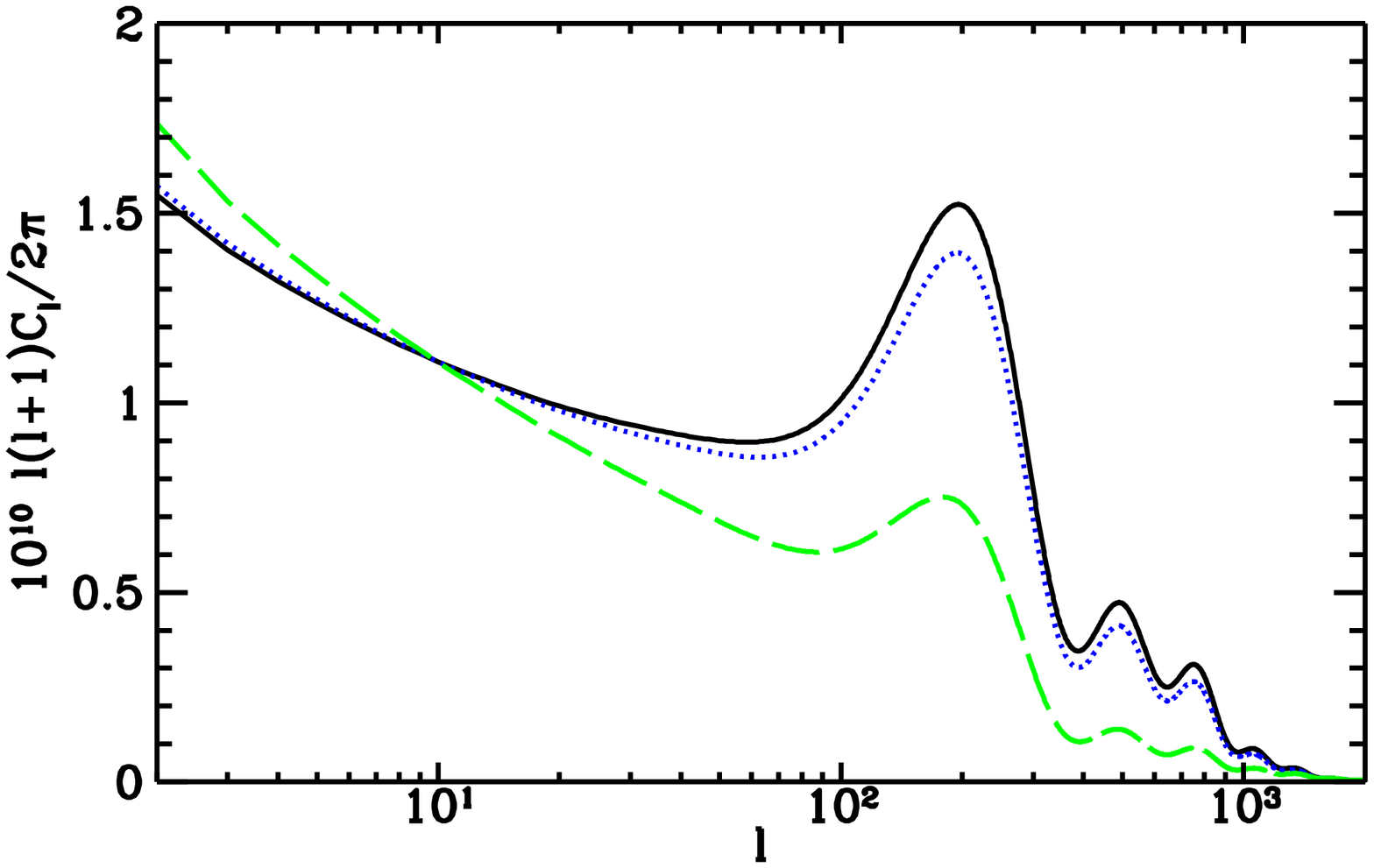}\\ 
\caption[fig4]{\label{cl} Predicted microwave anisotropies from the
power spectra computed under different assumptions summarized in Table
1: from top to bottom they are numerical, semi-analytic, and
lowest-order analytic, corresponding to the scalar and tensor spectra
shown in Fig.\ \ref{asat}.}
\end{figure}

Any set of observations will be able to probe only a range of scales,
and we shall use the capabilities of the Planck satellite as a
benchmark. Planck can probe the microwave anisotropy multipoles from
the quadrupole $\ell = 2$ up to around $\ell = 2000$. The
corresponding $k$ range sampled is from $k = a_0 H_0$ to $k = 1000 \,
a_0 H_0$. Using $a_0 H_0 = h/3000 \, {\rm Mpc}^{-1}$, the range of $k$
probed by Planck is shown in Fig.\ \ref{asat}.  We see that in fact
Planck would be able to see more or less to the bottom of the dip, but
not the subsequent rise. So the part of the spectrum to be probed does
not contain any very strong features. Fig.\ \ref{cl} shows the
predicted $C_\ell$ curves, both exact and approximate, calculated
using a modified version of the {\sc cmbfast} code \cite{cmbfast}. The
slight differences from the equivalent figure in Ref.~\cite{WMS} are
presumably from slightly different choices of parameters such as the
baryon density. The basic slow-roll method does extremely poorly.

To illustrate how to recognize when reconstruction is impossible, we
shall later consider the case where the scales left the horizon a
little further down the scalar field potential, corresponding to
shifting the spectrum leftwards by a factor of 30 and bringing the
main feature of the spectrum into the observable range.

\section{The Reconstruction Procedure}
\setcounter{equation}{0}

In this section we review the general reconstruction procedure, and
extend results for high derivatives of $V$ to higher-order in
slow-roll than previously given.  Perturbative reconstruction requires
that one fits an expansion, usually a Taylor series of the form
\begin{eqnarray}
\ln A_S^2(k) & = & \ln A_S^2(k_*) + (n_*-1) \ln \frac{k}{k_*} + \frac{1}{2}
	\, \left. \frac{dn}{d\ln k}\right|_* \ln^2\! \frac{k}{k_*}
	\nonumber \\
 & & \hspace*{-3em}  + \frac{1}{6} \left. \frac{d^2 n}{d (\ln k)^2} 
 	\right|_* \ln^3\! \frac{k}{k_*}  +\frac{1}{24} \left. 
 	\frac{d^3 n}{d (\ln k)^3} \right|_* \ln^4\! \frac{k}{k_*} + 
 	\cdots \,,
\end{eqnarray}
to the observed spectrum in order to extract the coefficients, where
stars indicate the value at $k_*$. The scale $k_*$ is most wisely
chosen to be at the (logarithmic) center of the data, so we take $k_*
= 0.01 \, h$ Mpc$^{-1}$.

How far the series should be taken is governed by how many of the
coefficients can be obtained with error bars inconsistent with
zero. Planck can in principle measure the $C_\ell$ at the percent
level of much of its range, though degeneracies with other parameters
will weaken the determination of the spectrum somewhat. This sets a
ballpark figure for how far the series should go, and for almost all
known inflation models only the amplitude and spectral index are
required.

In general one should similarly fit the tensor amplitude. The relative
amplitude of tensors and scalars is important for reconstruction. If
more information about the tensors can be determined than just the
amplitude, then the extra information is degenerate with information
in the scalars and the consistency relations can be tested, though it
seems extremely unlikely that observations will permit this
\cite{LLKCBA,parest,ZSS}.

{}From these, we reconstruct the potential, using the equations set
down in Ref.\ \cite{LLKCBA} plus similar equations for $V'''$ and
$V''''$.\footnote{For the potential we reconstruct the $V''''$ term is
negligible.}  This requires the exact relation connecting changes in
$\phi$ with changes in $k$ \cite{LLKCBA},
\begin{equation}
\label{phi-k}
\frac{d\phi}{d\ln k} = \frac{m_{{\rm Pl}}^2}{4\pi} \frac{H'}{H}
	\frac{1}{\epsilon-1} \simeq \sqrt{\frac{R_*}{4\pi}}\,(1+R_*)\,,
\end{equation}
The reconstruction equations, using the notation $n_*^{(i)}$ to
indicate $\left.  d^i n/d\ln k^i \right|_*$ and
$R_*=A_T^2(k_*)/A_S^2(k_*)$, are
\begin{eqnarray}
\label{vees}
m_{{\rm Pl}}^{-4}\, V(\phi_*) & \simeq & \frac{75} {32} A_S^2(k_*) R_* 
	\left\{ 1 +  \left[ 0.21 R_* \right] \, \right\} \nonumber \\ 
m_{{\rm Pl}}^{-3}\, V'(\phi_*) & \simeq & - \frac{75 \sqrt{\pi}}{8} 
	A_S^2(k_*) R_*^{3/2}\left\{ 1
	- \left[ 0.85R_*  -  0.53(1-n_*) \right] \,\right\}
	\nonumber \\
m_{{\rm Pl}}^{-2}\, V''(\phi_*) & \simeq & \frac{25\pi}{4}  A_S^2(k_*)  R_*
 	\left\{9 R_*  - 1.5(1-n_*)  \phantom{n_*^{(1)} } \right. \nonumber \\
 &  &\left. - \left[ 24.3 R_*^2  + 0.25 (1-n_*)^2 - 14.8  R_* (1-n_*) 
 	- 1.6 \, n_*^{(1)} \right]\, \right\}  \nonumber \\
m_{{\rm Pl}}^{-1}\, V'''(\phi_*) & \simeq & -\frac{75\pi^{3/2} }{4} A_S^2(k_*) R_*^{1/2}
	 \left\{ 24R_*^2 - 8R_* (1-n_*)  -  n_*^{(1)} \right. \nonumber \\
 & & + \left[    91 R_*^2 (1-n_*) - 9.2 R_* (1-n_*)^2  - 128R_*^3 \right.
                        \nonumber \\
 & &  \left. \left. + 12.3R_* \, n_*^{(1)}  +  0.03 (1-n_*) \, n_*^{(1)}  - 1.1 n_*^{(2)}
         \right]\,  \right\}    \nonumber \\
V''''(\phi_*) & \simeq & 75 \pi^2 A_S^2(k_*) \left\{ 60 R_*^3 
        - 30 R_*^2 (1-n_*) - 4.5 R_* \, n_*^{(1)}  \right. \nonumber \\
 & & +  2 R_* (1-n_*)^2 -  0.25 (1-n_*) \, n_*^{(1)} +  \ 0.5 \, n_*^{(2)}
 	\nonumber \\
 & &  + \left[ -524  R_*^4 +446R_*^3 (1-n_*)  + 0.55 \, n_*^{(3)}  
                   \right.  \nonumber \\
 & & +  0.02 (n_*^{(1)})^2  - 83.8R_*^2 (1-n_*)^2   +  65.7 R_*^2 \, 
n_*^{(1)} 
 	\nonumber \\
 & & +1.6 R_* (1-n_*)^3 - 6.8 R_*  \, n_*^{(2)} -  0.5 (1-n_*) \, n_*^{(2)}
 	\nonumber \\
 & & \left. \left. + 0.1 (1-n_*)^2 \, n_*^{(1)} -  11.1 R_*  (1-n_*) 
n_*^{(1)} 
 	\phantom{R_*^2} \hspace{-1em} \right] \right\}
\end{eqnarray}
We can assign a bookkeeping parameter to keep track of the order of
the terms appearing in Eq.\ (\ref{vees}).  $R_*$ and $1-n_*$ is of
order 1, and $n_*^{(j)}$ is of order $j+1$.  One can see that Eq.\
(\ref{vees}) is arranged so that the higher-order corrections appear
in square brackets.  Since the fit we use for the scalar power
spectrum only goes up to $n_*^{(2)}$, we have no information about
$n_*^{(3)}$, hence for consistency we only use the lowest-order
expression for $V''''$.

$R_*$, $n_*$, $n_*^{(1)}$, $\ldots$ are to be determined from
observations.  Fortunately, parameter estimation from the microwave
background has been explored in some detail \cite{parest,ZSS}. We
shall use error estimates for Planck assuming polarized detectors are
available, following the analysis of Zaldarriaga et al.\ \cite{ZSS}.
Most analyses have assumed that $R_*$ and $n_*$ are the only
parameters needed to describe the spectra.  In a related publication
\cite{CGL}, we have generalized their treatment to allow the power
spectrum to deviate from scale-invariance.  Including extra parameters
leads to a deterioration in the determination of {\em all} the
parameters, as it introduces extra parameter degeneracies.
Fortunately, for most parameters the uncertainty is not much increased
by including the first few derivatives of $n$ \cite{CGL}, but we have
found that the parameter $n$ itself has a greatly increased error
bar. If a power-law is assumed it can be determined to around $\Delta
n \simeq 0.004$ \cite{ZSS,CGL}, but including scale dependence
increases this error bar by a factor of ten or more. Notice that
unless one {\em assumes} a perfect power-law behaviour, this increase
in uncertainty is applicable even if the deviation from power-law
behaviour cannot be detected within the uncertainty.

{}From Ref.\ \cite{CGL}, an estimate of the relevant uncertainties is
\begin{eqnarray}
\label{obsun}
\Delta (R_*) \simeq 0.004 \; \; & ; & \; \; \Delta n \simeq 0.14 
\nonumber \\
\Delta (dn/d\ln k) \simeq 0.04 \; \; & ; & \; \; \Delta [d^2n/d(\ln k)^2 ]
	\simeq 0.005 \,.
\end{eqnarray}
Note that $r_{{\rm ts}}$ in Ref.\ \cite{CGL} is 14 times  
$R_*$.

\section{Two Reconstruction Examples}
\setcounter{equation}{0}

In this section we consider two reconstruction attempts of regions of the 
potential in Fig.\ 1. One is successful and the other unsuccessful. We 
demonstrate that this is to be expected and is perfectly consistent with the 
reconstruction procedure we have adopted, indicating that it is virtually 
impossible to misconstruct the potential.

\subsection{Successful reconstruction}
\setcounter{equation}{0}

In this subsection we reconstruct the potential shown in Fig.\
\ref{potential} that led to the spectra of Fig.\ \ref{asat}.  We find
that the observational errors dominate over the errors introduced by
using the slow-roll approximations for the spectra.  We do not agree
with claims in Ref.~\cite{WMS} that the breakdown of the slow-roll
approximation for this potential illustrates that reconstruction is
not valid for general potentials.  We also find that, as expected, the
shape of the potential is better determined than the amplitude.

\begin{figure}[!p]
\centering 
\leavevmode\epsfysize=8cm \epsfbox{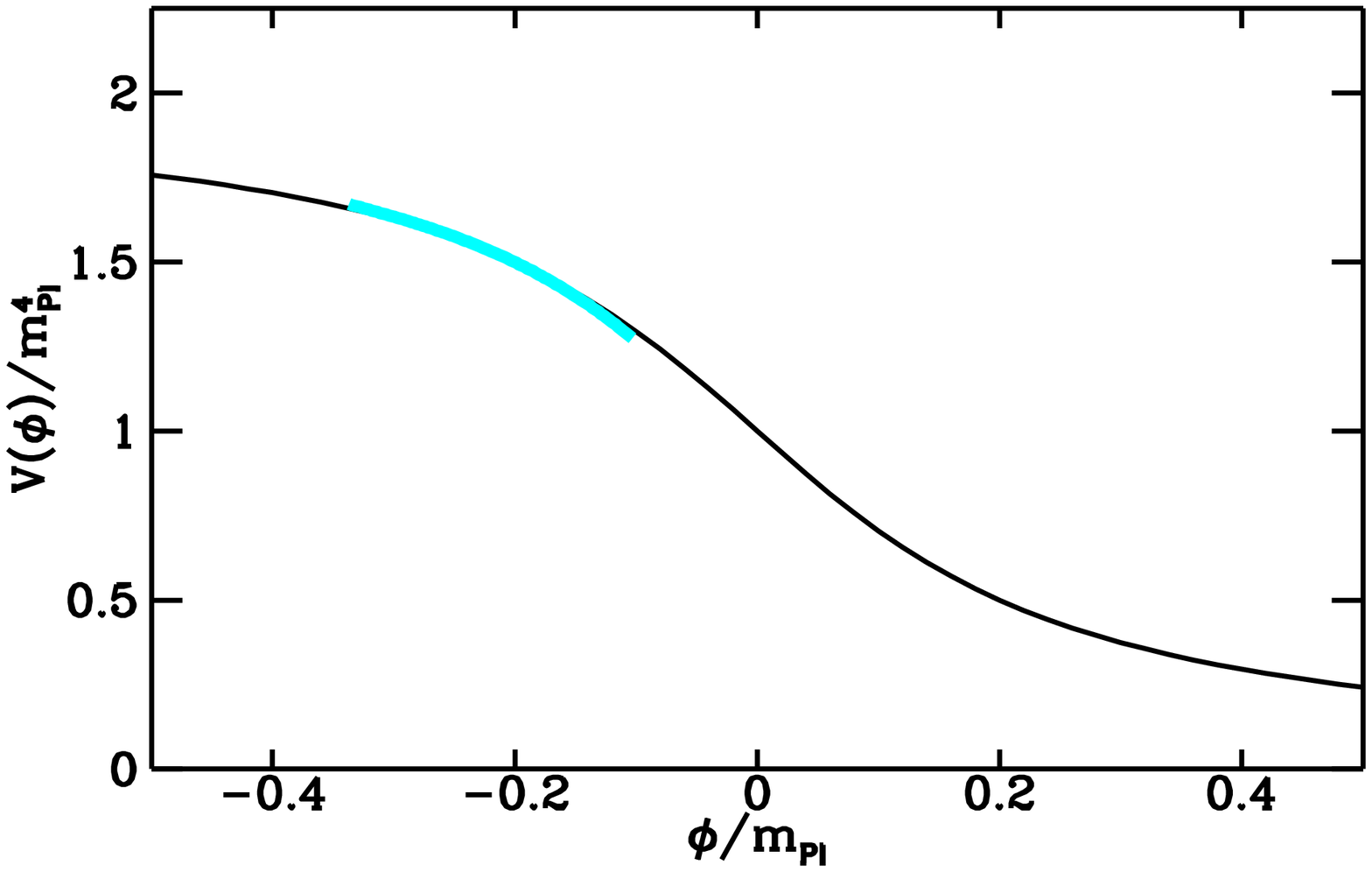}\\ 
\leavevmode\epsfysize=8cm \epsfbox{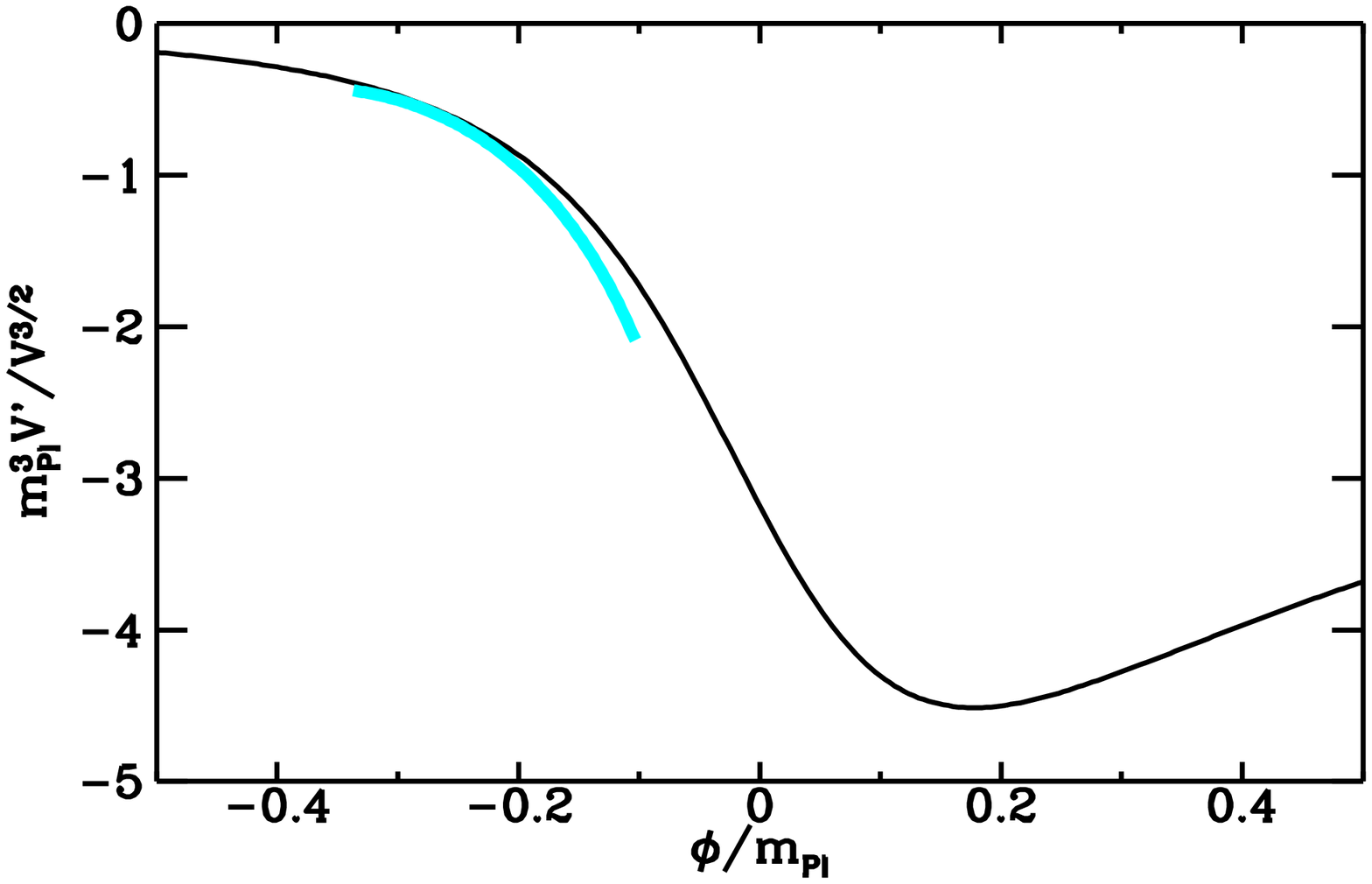}\\ 
\caption[fig5]{\label{v_theory} The original potential is compared
with the fragment reconstructed from Eq.\ (\ref{vees}) using the
parameter determination in Eq.\ (\ref{params}).  The lower figure is
the reconstructed $m_{Pl}^3V'/V^{3/2}$.  The difference between the
reconstructed fragments and the true potential is due to the use of
the slow-roll expansion in reconstruction. }
\end{figure} 

\begin{figure}[!p]
\centering 
\leavevmode\epsfysize=8cm \epsfbox{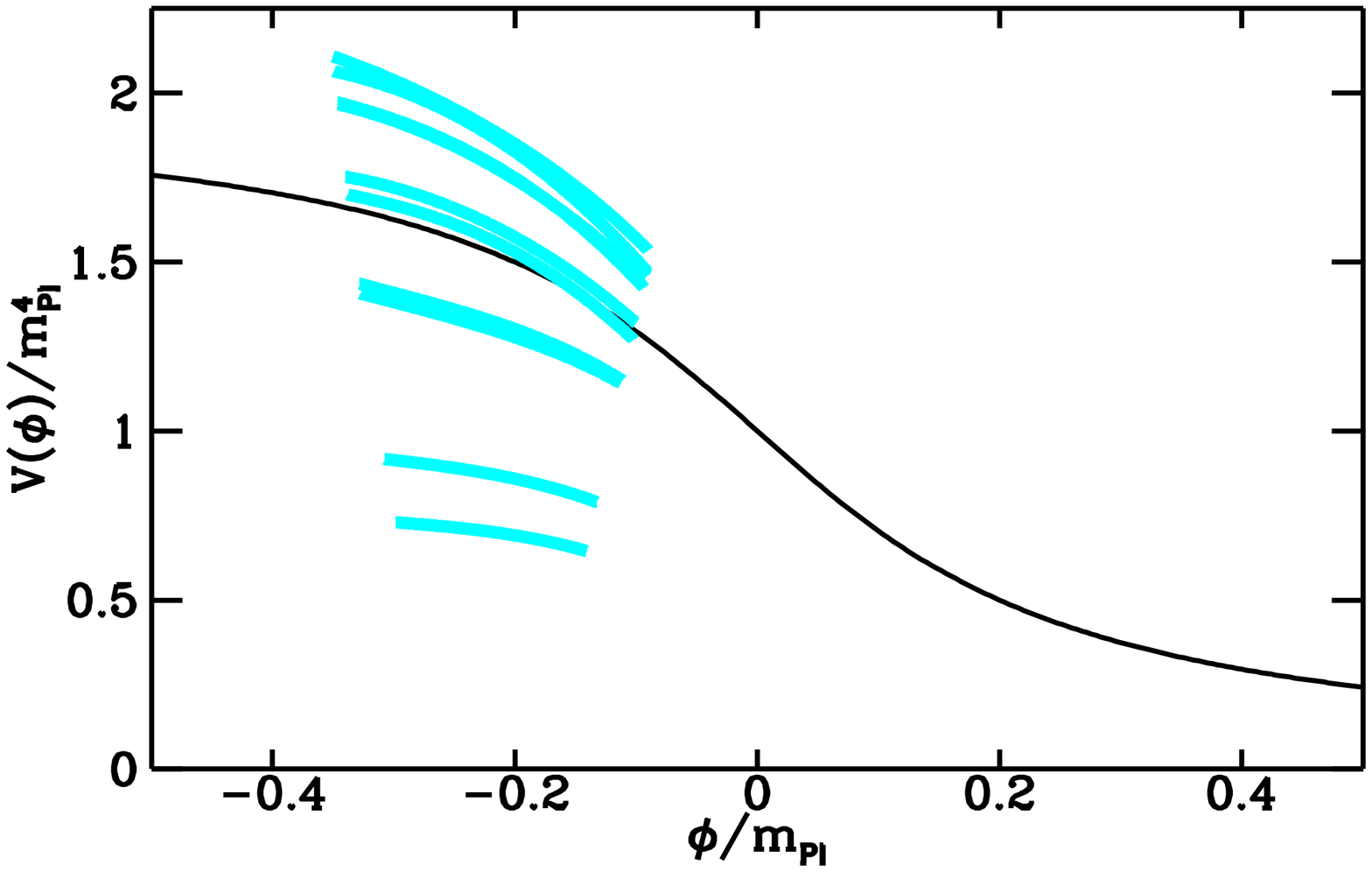}\\ 
\leavevmode\epsfysize=8cm \epsfbox{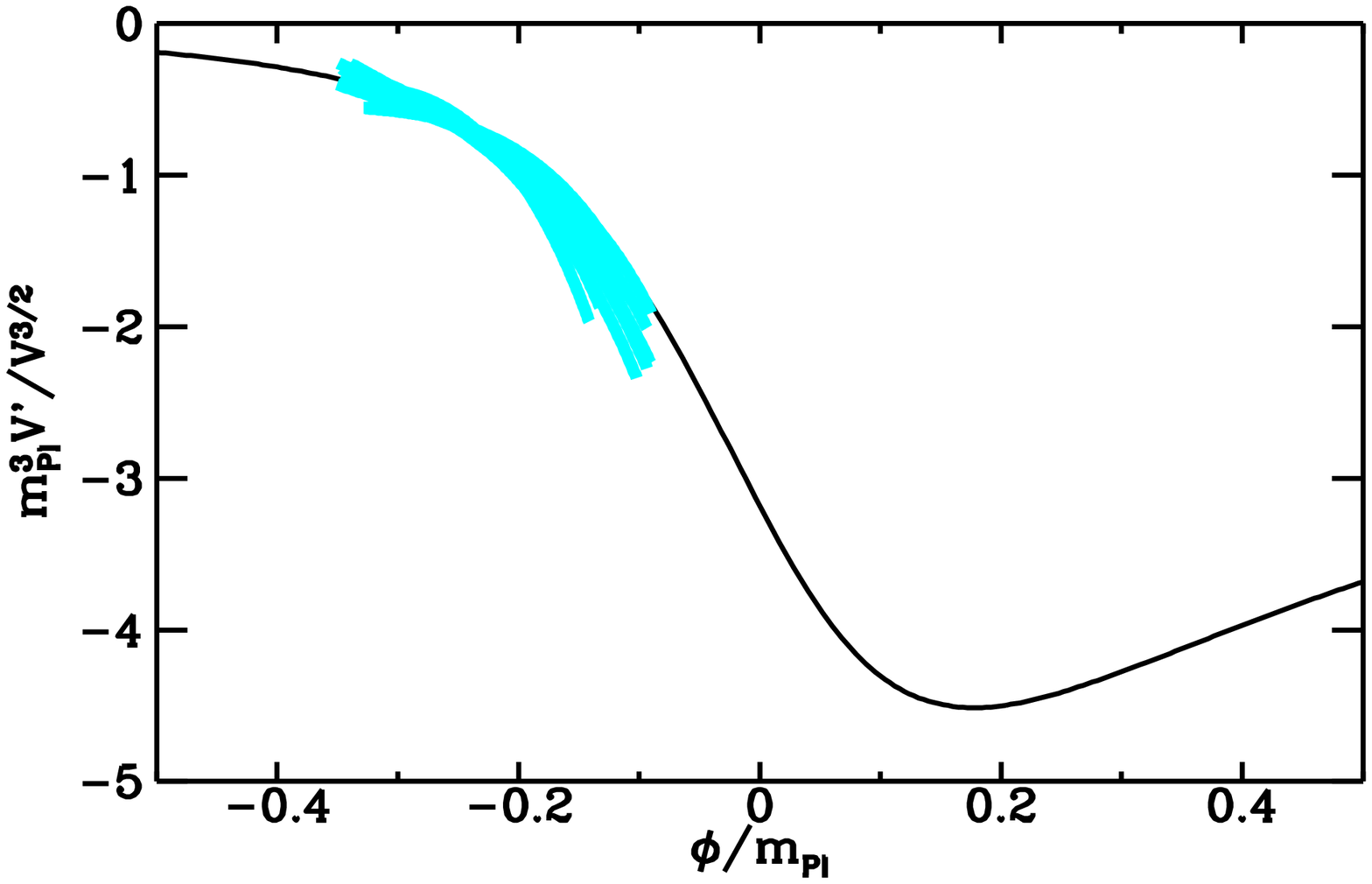}\\ 
\caption[fig6]{\label{v_obs} A set of ten reconstructed potential
fragments compared with the true potential.  The spread in the
reconstructed potentials compared to the reconstructed potential in
Fig.\ \ref{v_theory} illustrated our point that the uncertainty in the
parameter determination will overwhelm the error introduced by using
the slow-roll formalism in reconstruction.}
\end{figure} 

To estimate the errors introduced by the slow-roll formalism, we treat
the exact power spectra of Fig.\ \ref{asat} as the `observations', and
imagine that we do not know the potential from which they derive. We
shall later use estimated uncertainties assuming the spectra are
measured via the microwave anisotropies, but for now we shall imagine
that the power spectra have been measured exactly in order to assess
the errors arising from uncertainties in the theoretical power
spectrum determination.

We have fit a Taylor expansion of the above form to the exact
spectrum, using $k_*$ as above, and found the following results: 
\begin{eqnarray}
& & \ln A_S^2(k_*) = 3.84 \quad ; \quad R_*  =  0.014 \quad ; 
 \nonumber \\
& & n_*  =  0.579 \quad; \quad \left. \frac{dn}{d \ln k} \right|_* 
=  -0.134 \quad ; \quad
\left. \frac{d^2 n}{d (\ln k)^2} \right|_*  =   -0.052 \ .
\label{params}
\end{eqnarray}
{}From the estimated observational uncertainties of Eq.\
(\ref{obsun}), we see that all these coefficients are successfully
determined at high significance and a simple ``chi--by--eye''
demonstrates that the spectrum reconstructed from these data is an
adequate fit to the observed spectrum. This stresses the point that
the more unusual a potential is, the more information one is likely to
be able to extract about it, though the uncertainties on the
individual pieces of information may be greater.

We reconstruct the potential in a region 
about $\phi_* = -0.22$ (the reconstruction program does not
determine $\phi_*$), the width of the region given via Eq.\ (\ref{phi-k}). 
The results for the reconstructed potential {\em without} observational 
errors are shown in Fig.\ \ref{v_theory}, where it can immediately
be seen that the reconstruction has been very successful in
reproducing the main features of the potential while perturbations on
interesting scales are being developed.

Thus far we have assumed perfect observations, but of course in practice the 
measurement of the power spectrum is subject to observational errors. To 
illustrate visually the effect of the observational errors, we use a Monte 
Carlo procedure where we draw `observed' parameter values from gaussian 
distributions about the true value with the appropriate variance. These 
observed parameters, which include fully the effects of cosmic variance and 
the modelled instrument noise, are then fed into the reconstruction process. 
Fig.\ \ref{v_obs} shows ten such reconstructions about the true value. The 
uncertainty is indicated by the wide spread in the results. Note the width of 
the reconstructed region is correlated with the height of the potential; if 
the observed $R_*$ happens by chance to be low the potential is lower and 
flatter, so the field rolls more slowly as the perturbations are generated.

The uncertainty is dominated by that of $A_T^2(k_*)$; although the 
gravitational waves are detectable in this model, it is only about a 
three-sigma detection and the error bar is thus large. Since the overall 
magnitude of the potential is proportional to $A_T^2$, the visual impression 
is of a large uncertainty.

Fortunately, information in combinations of the higher derivatives is
more accurately determined. Fig.\ \ref{v_obs} also shows the
reconstruction of $V'/V^{3/2}$ with observational errors; this
combination is chosen as it is independent of the tensors to lowest
order.  Not only is it reconstructed well at the central point, but
both the gradient and curvature are well fit too, confirming useful
information has been obtained about not just $V''$ but $V'''$ as well,
which is only possible because of the extra information contained in
the scale-dependence of the power spectrum. So despite the poor visual
impression given by Fig.\ \ref{v_obs}, rather accurate information is
being obtained about the potential.

To compare these observational errors with the purely theoretical
errors of the slow-roll approximation, one can compare to Fig.\
\ref{v_theory}. From this we conclude that the error introduced by
using the bessel function approximation in reconstruction is a small
part of the error budget.

\subsection{An example where reconstruction is impossible}
\setcounter{equation}{0}

What we have seen in the example of Subsection 5.1 is that despite
slow-roll not working very well, and hence generating significant
errors in the normal estimation of the power spectrum, the
reconstructed potential still looks very good, especially in the light
of expected observational error bars. The main reason for this is that
although the spectrum is quite far from scale-invariance, it could
still be reasonably fit by just a few terms in the Taylor expansion.

The truly awkward situation would be a spectrum which could not be
adequately fit that way, for then perturbative reconstruction would
break down. This is no surprise---there is always a limit beyond which
a perturbative process cannot be taken. An example, alluded to above,
would be if inflation ended just a little later on the potential we
have been considering, so that the spectrum is shifted to the left
bringing the minimum into the observable range.

We have examined this case and found that indeed the spectrum cannot
be fit by a Taylor series to any reasonable order.  That
reconstruction fails is completely obvious, because we can't even
start. There is no danger of trying to carry out a reconstruction, and
obtaining an answer which has no connection to reality. If the
observations take this unfortunate turn, then one has a number of
options to try. Perhaps single-field inflation is not correct at all,
and some other theory such as topological defects or isocurvature fits
better. Or perhaps we do believe that a complicated inflaton potential
is at work, in which case a non-perturbative technique (e.g.\ fitting
the power in different wavebands\footnote{This technique was described
to us by Tarun Souradeep (private communication).}) would be more
appropriate. Or a multi-field inflation model might be responsible,
and one could test them on a model-by-model basis to see if they were
compatible with the data.

\section{Conclusions}
\setcounter{equation}{0}

We have found that a complicated potential, such as the example used
by Wang et al.\ \cite{WMS}, would be a boon to reconstruction of the
inflaton potential, as it would provide extra information accessible
to observations, in the form of scale-dependence of the density
perturbations.  Such potentials are also more likely to have tensor
modes at a detectable level, which is required for a complete
reconstruction to be performed.

We have analyzed the impact of observational errors on the
reconstruction, and conclude that any theoretical errors from use of
the slow-roll equations are likely to be sub-dominant. One might even
be able to test this be solving the mode equation for the
reconstructed potential, though we have not tried to pursue that route
here.

Finally, it has always been clear that there must be some limits to
the applicability of perturbative reconstruction, although it does a
good job even for the Wang et al.\ potential. However, such a failure
should be immediately evident from the data, as one would not be able
to successfully match observations with a truncated Taylor series
expansion of the power spectra. It would appear, therefore, that there
is no danger that perturbative reconstruction might appear to be
working, but in fact be producing an irrelevant potential.

\section*{Acknowledgments} 

EJC is supported by PPARC, IJG and ARL by the Royal Society and EWK by the 
DOE and NASA under grant NAG 5-2788. We thank Andrew Jaffe, Jim Lidsey and 
Tarun Souradeep for discussions. We acknowledge use of the Starlink computer 
system at the University of Sussex.


\end{document}